\documentclass[twocolumn,pre,showpacs,amsmath,longbibliography]{revtex4-1}
\usepackage[dvipdfmx]{graphicx}
\usepackage{url}
\usepackage{cases}
\usepackage{amssymb}
\usepackage{bm}
\usepackage{color}
\usepackage[version=3]{mhchem}
\usepackage[normalem]{ulem}
\usepackage{here}
\usepackage{comment}
\usepackage{physics}
\usepackage{siunitx} 

\begin{document}

\title{Phase coexistence in a weakly stochastic reaction-diffusion system}
\author{Yusuke Yanagisawa}
\affiliation{Department of Physics, Kyoto University, Kyoto 606-8502, Japan}
\author{Shin-ichi Sasa}
\affiliation{Department of Physics, Kyoto University, Kyoto 606-8502, Japan}
\date{\today}

\begin{abstract}
We investigate phase coexistence in a weakly stochastic reaction-diffusion
system without assuming a continuum description.
Concretely, for $(2N+1)$ diffusion-coupled vessels in which a
chemical reaction exhibiting bistability occurs, we derive a condition
for the phase coexistence in the limit $N \to \infty$. We then find that the
phase coexistence condition depends on the rate of hopping between neighboring
vessels. The conditions in the high- and low-hopping-rate limits are expressed
in terms of two different potentials which are determined from the chemical
reaction model in a single vessel.
\end{abstract}

\maketitle

{\em Introduction.---} 
Macroscopic systems out of equilibrium exhibit various ordered
structures called dissipative structures
\cite{Nicolis1977SelfOrganization,Kuramoto1984,cross1993pattern,beck1995thermodynamics}.
Although most of these studies are based on deterministic equations,
fluctuation effects on microscopic and mesoscopic systems out of equilibrium
have recently attracted much attention \cite{sekimoto2010, seifert2012stochastic}
because advances in experimental techniques have made it possible to observe, manipulate and control microscopic and mesoscopic dynamics 
\cite{moffitt2008recent,ciliberto2017experiments}. Examples include 
fluctuation effects on Turing patterns \cite{butler2009robust, biancalani2010stochastic, biancalani2017giant, falasco2018information, rana2020precision},
oscillation \cite{qian2002concentration}, chaos \cite{gaspard2020stochastic}, 
phase coexistence \cite{sasa2021stochastic}, and front propagation \cite{costantini2001asymmetric,iwata2010theoretical,khain2011fluctuations,kado2023microscopic}. 
In this Letter, we study fluctuation effects on phase coexistence in reaction-diffusion systems. 

In particular,  we focus on weakly stochastic systems. 
When there exists a unique stationary solution of the corresponding deterministic equation 
ignoring noise for the weakly stochastic system, 
the most probable value at the steady state is equal to the stationary solution.
In contrast, when multiple stationary solutions exist for the deterministic equation,
the most probable value would be determined 
as the stationary solution that maximizes the probability.
This has been fully understood through the study of simple stochastic chemical reaction models 
\cite{vellela2007quasistationary, endres2017entropy}. 

A similar but more complex situation occurs for reaction-diffusion systems. 
As an example, we consider a one-dimensional stochastic reaction-diffusion system  
where the chemical reaction has a bistable property.
For this system, there is a special parameter value
at which a spatially inhomogeneous stationary pattern appears. 
This pattern represents phase coexistence in one-dimensional space. 
When the system is described by a partial differential equation with weak noise, 
the condition for the phase coexistence is perturbatically obtained 
from that for the reaction-diffusion equation without noise \cite{costantini2001asymmetric, iwata2010theoretical, kado2023microscopic}.
However, when the interface width between the two phases is
much smaller than the coarse-graining length of the continuum
description, the validity of the continuum description is not ensured.

In this Letter, to study the phase coexistence condition for weakly stochastic reaction-diffusion systems 
without assuming a continuum description, 
we consider a stochastic process in $(2N+1)$ diffusion-coupled reaction vessels. 
We then find that the phase coexistence condition depends on the rate of hopping between the neighboring vessels.  
Remarkably, at the high-hopping-rate limit, the phase coexistence condition is obtained through the analysis of the reaction-diffusion equation, 
while at the low-hopping-rate limit, the phase coexistence condition is obtained through the perturbation analysis of the steady state. 
We show that these two conditions are expressed in terms of two different potentials, 
namely deterministic and stochastic potentials, which are determined from the chemical reaction model.

{\em Setup.---} 
First, we study a well-mixed chemical reaction system in a reaction vessel.
To be specific, we analyze the Schl\"{o}gl model \cite{schlogl1972chemical}
described by
\begin{align}
        \ce{\mathit{A} &<=>[k_{+1}][k_{-1}] \mathit{X}}, \\
        \ce{3\mathit{X} &<=>[k_{+2}][k_{-2}] 2\mathit{X} + \mathit{B}}.
\end{align}
Here, the concentrations of the chemical species $A$ and $B$ are fixed
and we consider the dynamics of the concentration of the chemical species $X$.
Let $\Omega$ be the volume of the system.
We use the same symbol $X$ to denote the number of particles of the chemical species $X$.
The concentrations of the chemical species $A,B$ and $X$ are denoted $a,b$ and $\xi$, respectively.
In the argument below, we set $a=1$ and $k_{+1}=0.5$ to make all quantities dimensionless.

We assume that the time evolution of $X$ is described by the Markov jump
process.
The probability distribution $P_t(X)$ of $X$ at time $t$ obeys
the master equation
\begin{align}
  \pdv{P_t(X)}{t} &=  W_+(X-1)P_t(X-1)+W_-(X+1)P_t(X+1)
                           \nonumber \\
                 &\quad -(W_+(X)+W_-(X))P_t(X),
    \label{master_eq}
\end{align}
where the transition rates $W_+$ and $W_-$ for
the Schl\"{o}gl model are given by
\begin{align}
    W_+(X) &= k_{+1}a\Omega + k_{-2}bX(X-1)/\Omega, \\
    W_-(X) &= k_{-1}X + k_{+2}X(X-1)(X-2)/\Omega^2, 
\end{align}
respectively.
The stationary solution of the master equation takes the asymptotic form
\begin{align}
    P_{\rm ss}(X) = \exp(-\Omega V_{\mathrm{sto}}(\xi)+o(\Omega))
    \label{master_ss_sol}
\end{align}
for large $\Omega$.
We refer to $V_{\mathrm{sto}}(\xi)$ as a stochastic potential.
We calculate $V_{\mathrm{sto}}(\xi)$ according to
\begin{align}
  \dv{V_{\mathrm{sto}}(\xi)}{\xi}
  = \ln \left( \frac{w_-(\xi)}{w_+(\xi)} \right)
\label{def:spot}  
\end{align}
with normalized transition rates
\begin{align}
    w_+(\xi) &= k_{+1}a + k_{-2}b\xi^2, \\
    w_-(\xi) &= k_{-1}\xi + k_{+2}\xi^3.
\end{align}
As stochastic effects on the time evolution of the
concentration $\xi$ become weaker for larger $\Omega$ \cite{gardiner1985handbook}, 
a system with large $\Omega$ is identified as a weakly stochastic system, 
which is the target of this Letter.

\begin{figure}
    \centering
    \includegraphics[width=0.9\linewidth]{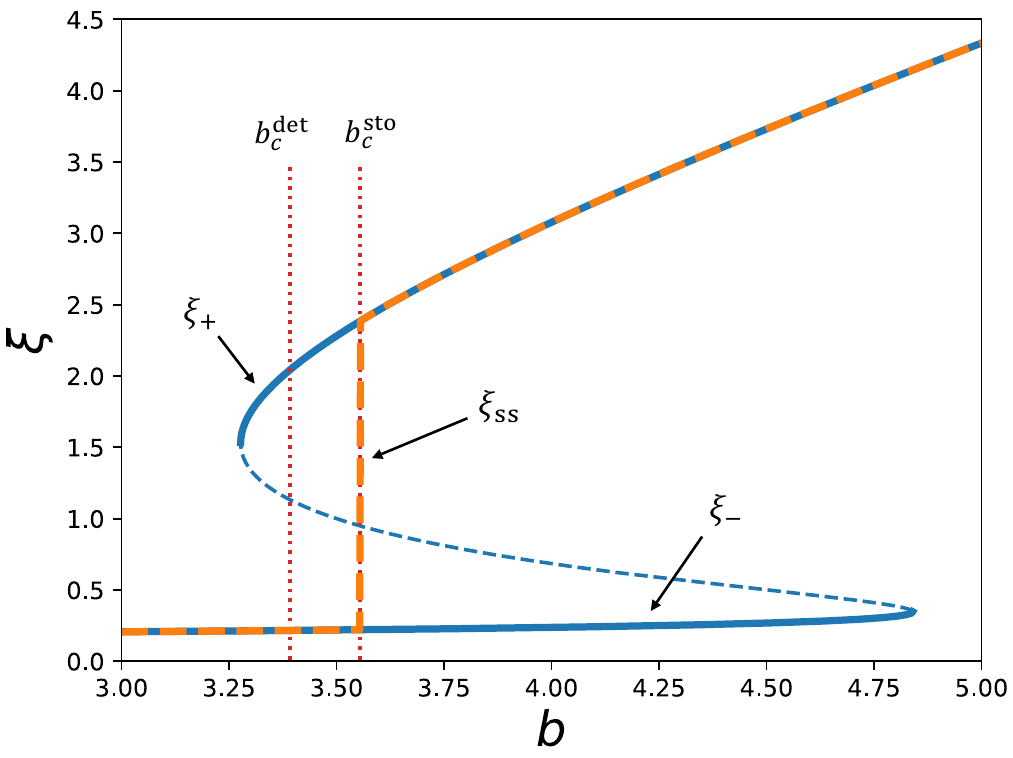}
    \caption{Relationship between $\xi_+, \xi_-$ and $\xi_{\mathrm{ss}}$.
    In the weakly stochastic system, $\xi_{\mathrm{ss}}$ is equal to either $\xi_+$ or $\xi_-$, and it is determined
    as the minimum point of the stochastic potential $V_{\mathrm{sto}}(\xi)$ for each $b$.}
    \label{bifurcation_diagram}
\end{figure}

The stationary concentration $\xi_{\mathrm{ss}}$ of the weakly stochastic system 
is defined as the expectation value of $\xi$ by first taking the limit $t\to\infty$ in \eqref{master_eq},
and then considering the limit $\Omega\to\infty$ in \eqref{master_ss_sol}.
Note that $\xi_{\mathrm{ss}}$ minimizes the stochastic potential
$V_{\mathrm{sto}}(\xi)$ given by \eqref{def:spot}. 
In contrast, one can also consider the case of first taking the
limit $\Omega\to\infty$ in \eqref{master_eq}. 
The concentration dynamics then follows the deterministic rate equation
described by
\begin{align}
    \dv{\xi}{t} &= -k_{+2}\xi^3 +k_{-2}b\xi^2 -k_{-1}\xi +k_{+1}a.
    \label{rate_eq}
\end{align}
The right-hand side of the rate equation is 
written as a potential force $-dV_{\mathrm{det}}/d\xi$, where
\begin{align}
    V_{\mathrm{det}}(\xi) = \frac{1}{4}k_{+2}\xi^4-\frac{1}{3}k_{-2}b\xi^3+\frac{1}{2}k_{-1}\xi^2-k_{+1}a\xi.
    \label{det_ptl}
\end{align}
We refer to $V_{\mathrm{det}}(\xi)$ as a deterministic potential.
Using \eqref{det_ptl}, we find that the rate equation with some parameter values has two stable stationary solutions $\xi_-$ and $\xi_+$, $\xi_- < \xi_+$,
at which $V_{\mathrm{det}}(\xi)$ becomes local minima. Note that 
the stationary concentration $\xi_{\mathrm{ss}}$ is equal to either $\xi_-$ or $\xi_+$ as shown in Fig. \ref{bifurcation_diagram} \cite{endres2017entropy}.

Each of the two potentials $V_{\mathrm{sto}}$ and $V_{\mathrm{det}}$ has two local minima 
for some parameter values.
Below, we focus on this case with changing $b$ while the other parameters are fixed. 
We define $b_c^{\mathrm{sto}}$ as the value 
at which the two local minimum values of $V_{\mathrm{sto}}$ coincide. 
Similarly, we can define $b_c^{\mathrm{det}}$ from $V_{\mathrm{det}}$.
Note that, in general, $b_c^{\mathrm{sto}}\neq b_c^{\mathrm{det}}$. 
For later convenience, let $\kappa_{\pm}$ be the rate of relaxation to $\xi_{\pm}$, 
which provides the characteristic time of the chemical reaction. 
Explicitly, we have $\kappa_\pm \equiv |V_{\mathrm{det}}''(\xi_\pm)|
=k_{+2}(\xi_\pm-\xi_0)(\xi_\pm-\xi_\mp)$, 
where $\xi_0$ is the local maximum point of $V_{\mathrm{det}}(\xi)$.

\begin{figure}
    \centering
    \includegraphics[width=0.98\linewidth]{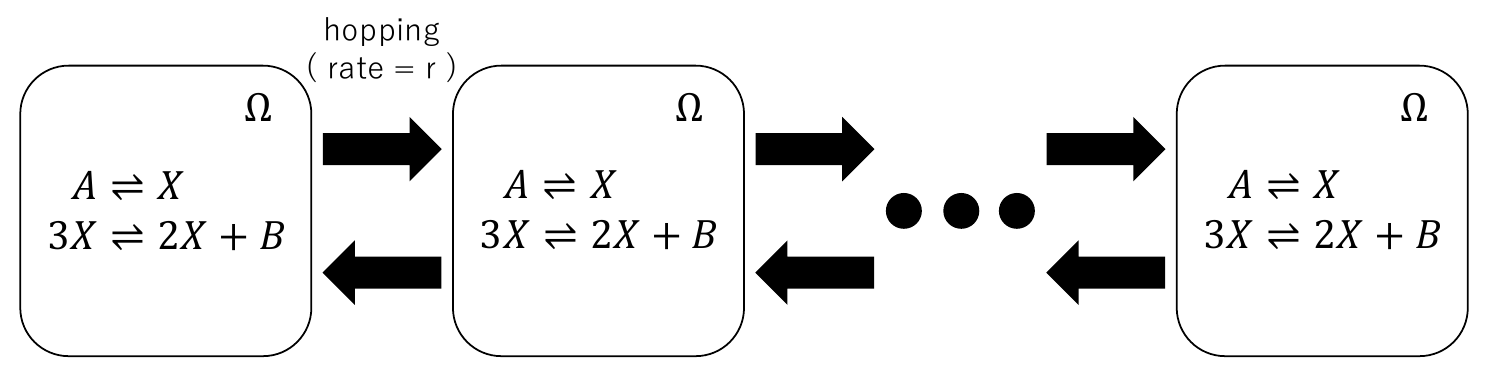}
    \caption{Schematic figure of a stochastic reaction-diffusion system.}
    \label{Fig_stochastic_RD}
\end{figure}

We now introduce a stochastic reaction-diffusion system \cite{erban2020stochastic}.
As shown in Fig. \ref{Fig_stochastic_RD},
$2N+1$ reaction vessels are arranged in the $x$ direction.
Let $\Delta x$ be the length of each vessel in the $x$ direction and $\mathcal{S}$ be the area of the cross-section perpendicular to the $x$ direction.
The volume of each vessel is  given by $\Omega\equiv \mathcal{S}\Delta x$.
The numbers of particles $A$ and $B$ are fixed as the same values for all vessels unless otherwise mentioned later.
Each vessel is assumed to be well mixed, and the state of the system is described by $\vb*{X}\equiv (X_{-N}, \cdots, X_N)$, a collection of the particle numbers of $X$. 
Here, we assume that chemical reactions occur only between particles in the same vessel 
and that particles hop to adjacent vessels.
All chemical reactions and particle hoppings are described by the following Markov jump processes.
First, state transitions associated with chemical reactions in the $i$-th vessel, $X_i \to X_i\pm 1$,
occur at the transition rate $W_\pm(X_i)$.
Second, state transitions associated with particle hoppings starting from the $i$-th vessel,
$(X_i, X_{i+\sigma}) \to (X_i-1, X_{i+\sigma}+1)$ with $\sigma \in \{1, -1\}$,
occur at the transition rate $rX_i$.
The constant $r$ represents the hopping rate per one particle.
Note that particles in the vessels at either end hop inward only.

The aim of this Letter is to derive the phase coexistence condition
for the weakly stochastic reaction-diffusion system in the limit $N\to\infty$.
Here, a phase coexistence state is defined as the spatially
inhomogeneous steady state connecting two locally stable stationary concentrations.
In particular, we consider two limiting cases, 
namely (i) the high-hopping-rate regime in which $r/\kappa_{\pm} \gg 1$
and (ii) the low-hopping-rate regime in which $r/\kappa_{\pm} \ll 1$.
For both regimes, we first take $N\to\infty$ and then $t\to\infty$ 
and we finally take $\Omega\to\infty$.

Before studying the two limiting cases,
we consider the limit $\Omega \to \infty$
in the weakly stochastic reaction-diffusion system.
We define $\phi_i(t)\equiv \langle X_i\rangle_t/\Omega$,
where $\langle\cdots\rangle_t$ represents the expectation for the 
stochastic process. Then, in the limit $\Omega\to\infty$, the time
evolution of $\phi_i(t)$ is described by \cite{erban2020stochastic, Note10}
\begin{align}
    \partial_t\phi_i &= -k_{+2}\phi_i^3+k_{-2}b\phi_i^2-k_{-1}\phi_i+k_{+1}a
    \nonumber \\
    &\quad + r (\phi_{i+1}-2\phi_i+\phi_{i-1}).
    \label{descrete_phi}
\end{align}
Here, we consider the stationary solution of the deterministic equation
by taking the limit $t \to \infty$. If the stationary solution 
is uniquely determined, the order of the two limits $\Omega \to \infty$ and $t \to \infty$
can be exchanged. Therefore, to determine the phase coexistence condition
in the limit $N \to \infty$, we first investigate the uniqueness of the
stationary solution.

{\em High hopping rate regime.---}
For the diffusion constant defined by $D \equiv r(\Delta x)^2$, 
the length reached by the diffusion during the characteristic
chemical reaction time $\kappa_{\pm}^{-1}$ is given by
$\ell^d_\pm\equiv \sqrt{D/\kappa_{\pm}}$ \cite{kuramoto1974effects}. 
The length is referred to as the Kuramoto length \cite{van1992stochastic}. 
Then, the condition $r \gg \kappa_{\pm}$ implies $\ell^d_{\pm} \gg \Delta x$. 
In this case, the continuum description
using $\phi(x,t)\equiv \phi_i(t)$ with $x=i \Delta x$ becomes the reaction-diffusion equation 
\begin{align}
    \partial_t \phi = -k_{+2}\phi^3+k_{-2}b\phi^2-k_{-1}\phi+k_{+1}a + D\partial_x^2 \phi,
    \label{RD_eq}
\end{align}
which  provides an asymptotically correct prediction for the solution of \eqref{descrete_phi}.
Using $L\equiv N \Delta x$, $\phi(x,t)$ is defined 
in a one-dimensional space $x\in [-L,L]$. 
As $N \to \infty$ is equal to $L/\ell^d_\pm \to \infty$, 
phase coexistence states are expressed by the stationary solution
satisfying the boundary conditions
$\phi(x=-L) = \xi_-$ and $\phi(x=L) = \xi_+$ in the limit $L/\ell^d_\pm \to \infty$. 
We find the unique stationary solution in this case, 
and we thus conjecture reasonably that 
\eqref{descrete_phi} possesses the unique stationary solution under the boundary conditions
$\lim_{N  \to \infty} \phi_{-N} = \xi_-$ and $\lim_{N \to \infty} \phi_N = \xi_+$
when $r/\kappa_{\pm} \gg 1$. We thus exchange the order of
the limits $\Omega \to \infty$ and $t \to \infty$ and conclude that the phase
coexistence condition is given by the point where the deterministic potential
$V_{\mathrm{det}}(\xi)$ takes the same value at the two minima \cite{Note10}.
Therefore, $b=b_c^{\mathrm{det}}$ is the phase coexistence condition
for case (i).

\begin{figure}
    \centering
    \includegraphics[width=0.95\linewidth]{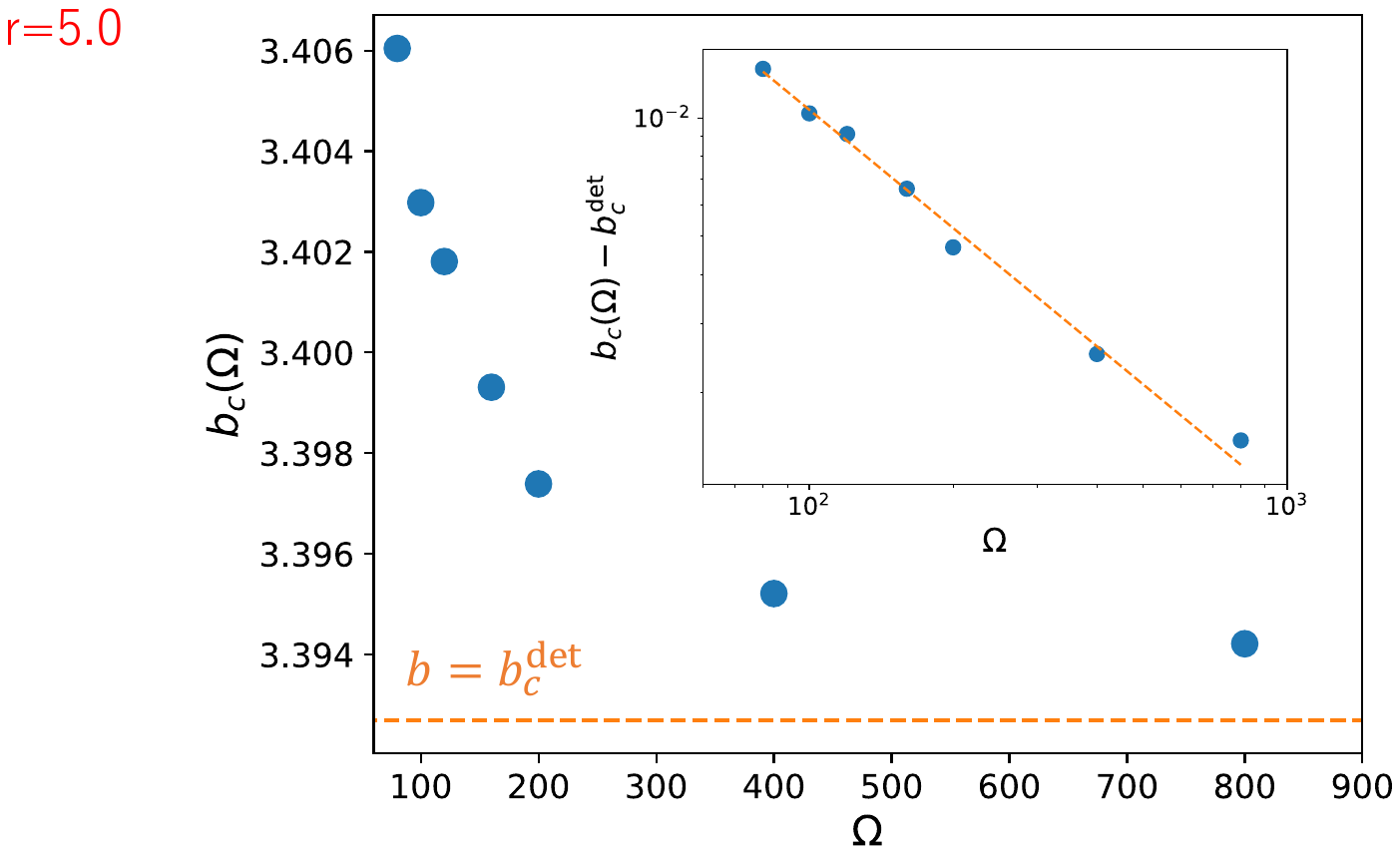}
    \caption{$\Omega$ dependence of $b_c(\Omega)$ for $r=5.0$. The inset
    shows a log-log plot with the guide line $b_c(\Omega)-b_c^{\mathrm{det}} = \alpha/\Omega$, where $\alpha=1.05$.}
    \label{bc-Omega_relation}
\end{figure}

We confirm the conjecture through direct numerical simulations of
the stochastic reaction-diffusion system.
As an example, chemical reaction constants are assumed as
$k_{-1}=3.0$, and $k_{+2}=k_{-2}=1.0$, 
where $\kappa_+=\kappa_-=1.67$ at $b=b^{\rm det}_c$. 
To satisfy the condition $r \gg \kappa_{\pm}$, we choose $r=5.0$. Other parameter
values are set to $N=75$ and $\Delta x=1.0$.
As an initial condition, $X_i(0)/\Omega = \xi_-$ for $i<0$, 
$X_i(0)/\Omega = \xi_+$ for $i > 0$, and $X_0(0)/\Omega=\xi_0$.
To numerically determine the phase coexistence condition, we measure 
the interface position $\hat{z}(t)$ at time $t$,
where $\hat{z}(t)$ is given by the $x$-coordinate
at which the concentration profile $\vb*{X}(t)$ crosses the reference concentration $(\xi_++\xi_-)/2$.
We then estimate the interface velocity $v(b)$ as the statistical average of
the displacement of the interface between $t=t_0$ and $t=t_1$:
\begin{align}
    v(b) = \frac{\langle \hat{z}(t_1)-\hat{z}(t_0) \rangle}{t_1-t_0},
\end{align}
where $t_0$ is much larger than the relaxation time $\kappa_{\pm}^{-1}$
and $t_1-t_0$ is large enough to detect the average velocity.
Specifically, we set $t_0=10$ and $t_1=100$, and
the expectation value is estimated by making $100$
independent calculations of $\hat z(t)$. 
The phase coexistence condition $b=b_c(\Omega)$ for the system with finite $\Omega$ is determined by $v(b_c(\Omega))=0$.
Figure \ref{bc-Omega_relation} shows the $\Omega$ dependence of $b_c(\Omega)$.
It is observed that $b_c(\Omega)$ approaches $b_c^{\mathrm{det}}$ as $\Omega$ increases. 
Such a difference between $b_c(\Omega)$ and $b_c^{\mathrm{det}}$ was reported in Ref. \cite{khain2011fluctuations}.
More quantitatively, we numerically find that
$b_c(\Omega)-b_c^{\mathrm{det}} \simeq 1/\Omega$. Similar behaviors are
observed for $r=0.5$ and $r=1.0$ \cite{Note10}. 
This leads to a conjecture that $b_c(\Omega) \to b_c^{\mathrm{det}}$ as
$\Omega \to \infty$ for $r$ in a certain range $[r_c,\infty]$. 
In addition, the numerical result suggests that $r_c$ is smaller than $\kappa_{\pm}$, 
which is not understood from a theoretical viewpoint.

{\em Low hopping rate regime.---}
Next, we consider the low-hopping-rate regime satisfying $r \ll \kappa_\pm$.
It has been shown that, for a model similar to ours, an equation corresponding to \eqref{descrete_phi} has many stationary solutions in the low-hopping-rate regime \cite{berglund2007metastability}.
Therefore, to determine the phase coexistence condition, we first have to consider the limit $t\to\infty$; i.e., we have to deal with the original stochastic process.
We thus focus on the stationary solution of the master equation with large $\Omega$. 
This solution takes the asymptotic form 
\begin{align}
  P_{\mathrm{ss}}(\vb*{X})
  = \exp(-\Omega\tilde{V}_{\mathrm{sto}}(\vb*{\xi};b)+o(\Omega)),
    \label{ss_solution}
\end{align}
where $\vb*{\xi}=(\xi_{-N}, \cdots, \xi_N)$.
The most probable configuration minimizes the potential
$\tilde{V}_{\mathrm{sto}}(\vb*{\xi};b)$.
When this configuration is not homogeneous in space for a certain value of the
parameter $b$, we identify the phase coexistence at the point
$b=b_c^{\mathrm{low}}$.

To extract the phase coexistence condition in the limit $N\to\infty$, we consider a ramped system in which
\begin{align}
    b_i = b + \frac{i}{N}\Delta b \equiv b + \delta b_i
    \label{b_ramping}
\end{align}
is defined for the $i$-th vessel.
The phase coexistence is more likely to occur in the ramped system than in the original system.
The phase coexistence condition in the original system can be obtained by considering the limit $N\to\infty$ and $\Delta b\to 0$.

Noting that the hopping rate $r$ is small, we formulate a perturbation theory
for calculating the potential $\tilde{V}_{\mathrm{sto}}(\vb*{\xi};b, \Delta b)$
in the ramped system. 
Ignoring $o(r)$ terms and $o(\Delta b)$ terms in the limit $r \to 0$ and $\Delta b \to 0$, 
we write the leading order form of the potential as 
\begin{align}
\tilde{V}_{\mathrm{sto}}(\vb*{\xi};b,\Delta b)
&= \sum_{i=-N}^N V_{\mathrm{sto}}(\xi_i;b) \nonumber \\
&\quad  +\sum_{i=-N}^N \partial_b V_{\mathrm{sto}}(\xi_i;b)\cdot\delta b_i  
        +rJ(\vb*{\xi}),  
\label{V_expansion}
\end{align}
where $V_{\mathrm{sto}}(\xi_i;b)$ is the stochastic potential given by
\eqref{def:spot} explicitly indicating the $b$ dependence. 
Substituting \eqref{V_expansion} into the master equation of the ramped system, 
we obtain the equation for $J(\vb*{\xi})$ as \cite{Note10}
\begin{align}
 &\sum_{i=-N}^N (w_+(\xi_i)-w_-(\xi_i))\partial_{\xi_i}J \nonumber \\
=& \sum_{i=-N}^{N-1}
   [\xi_i (g_i(\vb*{\xi})-1) 
   +\xi_{i+1} (g_i(\vb*{\xi})^{-1}-1) ]
    \label{J_PDE}
\end{align}
with $g_i(\vb*{\xi})= w_+(\xi_i) w_-(\xi_{i+1})/[w_-(\xi_i)w_+(\xi_{i+1})]$.
By numerically solving  \eqref{J_PDE}, we obtain $J(\vb*{\xi})$. 
As a result, $\tilde{V}_{\mathrm{sto}}(\vb*{\xi};b,\Delta b)$
is obtained for each configuration $\vb*{\xi}$ \cite{Note10}.  

\begin{figure}
    \centering
    \includegraphics[width=0.9\linewidth]{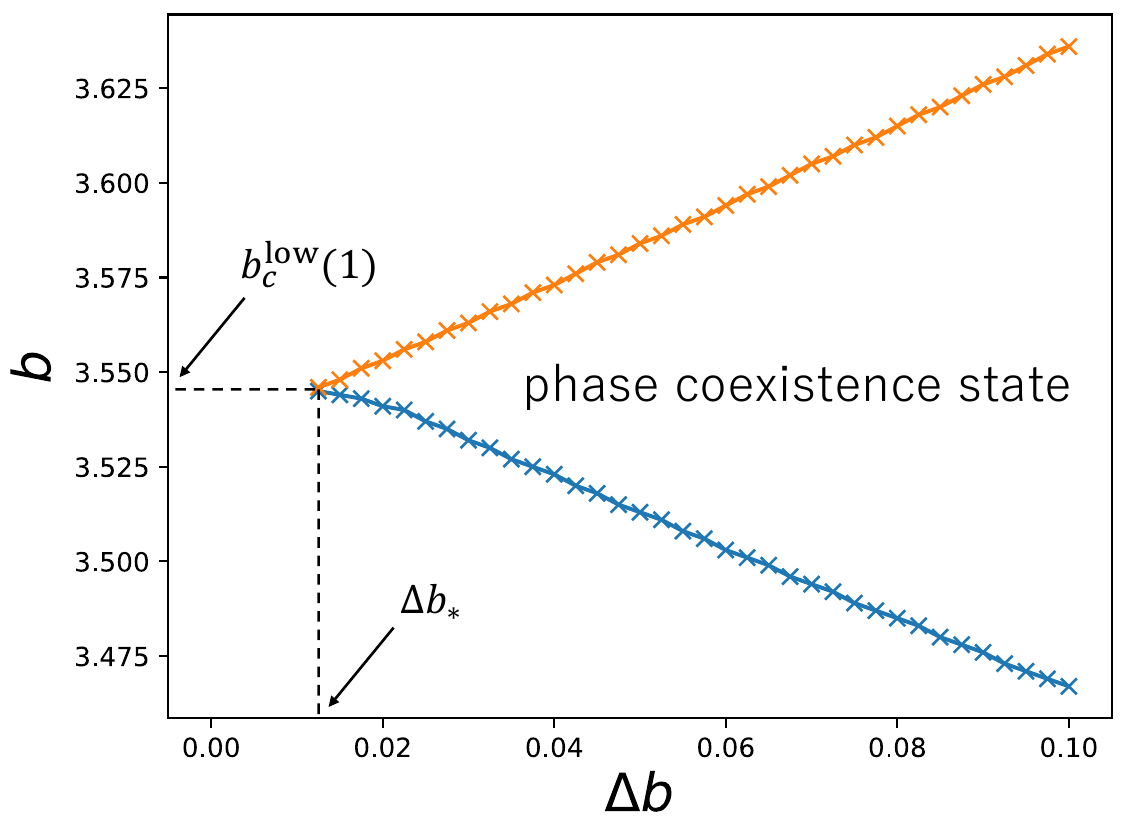}
    \caption{Phase diagram in the $(b, \Delta b)$ space for $N=1$.}
    \label{b-delb_relation}
\end{figure}

To find the configuration that minimizes
$\tilde{V}_{\mathrm{sto}}(\vb*{\xi};b,\Delta b)$, 
we consider $2N+2$ configurations $\vb*{\xi}^{(0)},\vb*{\xi}^{(1)},\cdots,\vb*{\xi}^{(2N+1)}$
as candidates of the minimizer, 
where for any integer $k$ satisfying $0\leq k \leq 2N+1$,
$\xi^{(k)}_i=\xi_-$ for $-N \le i \le N-k$ and $\xi^{(k)}_i=\xi_+$ for $N-k+1 \le i \le N$. 
Let $k_*$ be an integer such that $\vb*{\xi}^{(k_*)}$ provides the minimum value of 
$\tilde{V}_{\mathrm{sto}}(\vb*{\xi}^{(k)};b,\Delta b)$ for a given $(b, \Delta b)$. 
If $k_* \neq 0,\ 2N+1$, an inhomogeneous configuration
$\vb*{\xi}^{(k_*)}$ is most probable for the ramped system with $b$ and $\Delta b$.
In this case, we identify  the phase coexistence state.
Figure \ref{b-delb_relation} is a phase diagram in the parameter space $(b,\Delta b)$ for $N=1$. 
Similar diagrams can be obtained for any finite $N$.
The diagrams show that a positive $\Delta b_*$ is necessary to observe
the phase coexistence state for systems with finite $N$.
The phase coexistence condition for the special value $\Delta b_*$
is denoted by $b_c^{\mathrm{low}}(N)$. We then find that 
$\Delta b_*$ approaches $0$ following the law $\Delta b_*\simeq 1/N$ \cite{Note10}.
Therefore, in the limit $N \to\infty$, the phase coexistence
state can be realized at $b=b_c^{\mathrm{low}}(\infty)$ without ramping.
This provides the phase coexistence condition for the original stochastic reaction-diffusion system in  the low-hopping-rate regime.

\begin{figure}
    \centering
    \includegraphics[width=0.9\linewidth]{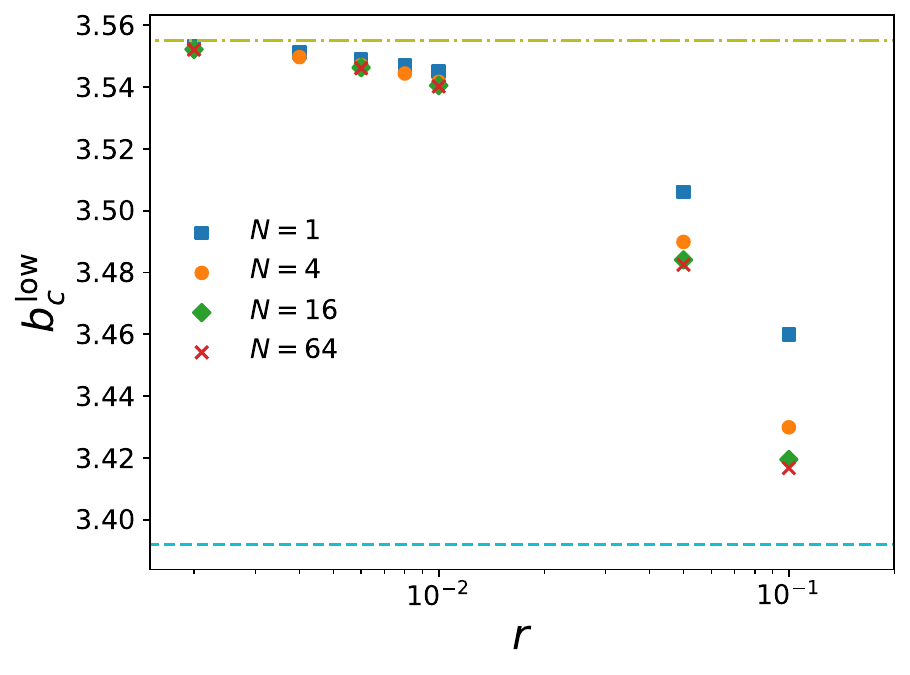}
    \caption{Hopping rate dependence of the phase coexistence condition $b_c^{\mathrm{low}}(N)$ in the low-hopping-rate regime.
      In the limit $N\to\infty$, $b_c^{\mathrm{low}}(N)$ converges to a certain value that is clearly different from $b_c^{\mathrm{det}}$ shown by the blue dashed
      line. The yellow dash-dot line shows the value of $b_c^{\mathrm{sto}}$.}
    \label{bc-r_relation}
\end{figure}

In Fig. \ref{bc-r_relation}, $b_c^{\mathrm{low}}(N)$ is presented
as a function of $r$ for several values of $N$.
It is observed that $b_c^{\mathrm{low}}(N)$ for each $r$ converges to a certain
value in the limit $N \to \infty$. 
Interestingly, the convergence value $b_c^{\mathrm{low}}(\infty)$ is clearly different from the phase coexistence condition $b_c^{\mathrm{det}}$ derived from the reaction-diffusion equation. 
Moreover, we find that $b_c^{\mathrm{low}}(\infty)$ takes the value around $b_c^{\mathrm{sto}}$. 
This means that the results in the low-hopping-rate regime
are qualitatively different from that in the high-hopping-rate regime
and suggests that a phase transition occurs at some point $r_c$
beyond which the phase coexistence condition is described by the reaction-diffusion equation.

{\em Concluding Remarks.---}
In this Letter, we calculated the phase coexistence condition
for the weakly stochastic reaction-diffusion system.
We found that the phase coexistence condition depends on the hopping rate. 
In the high-hopping-rate regime, the phase coexistence is described 
by the deterministic reaction-diffusion equation
and thus determined by the deterministic potential. 
In contrast, the phase coexistence in the low-hopping-rate regime
is determined by a perturbation from the stochastic potential.  

To the best of our knowledge, the latter type of phase coexistence has never been observed in reaction-diffusion systems
as these systems have typically been studied using macroscopic setups \cite{kondo2010reaction}. 
One possible way to observe such phase coexistence is to use an assemble of bacteria. 
One can consider a bistable chemical reaction in a cell of each bacterium. 
If the chemical species are controlled to flow out of the cell, 
the diffusion-coupled chemical reaction can be designed by preparing many bacteria. 
Indeed, such a system has been studied for stochastic pattern formation \cite{karig2018stochastic}. 
It would be interesting to observe phase coexistence 
in the low-hopping-rate regime \cite{Note10}. 

From the theoretical viewpoint, the most important problem is
to determine whether a phase transition occurs as a function of $r$.
To solve the problem, we need to study the phase coexistence condition for any hopping rate $r$. 
As one approach, one can study the Hamilton-Jacobi equation for determining the most probable configuration, 
which is a natural generalization of our formulation \eqref{J_PDE} \cite{kubo1973fluctuation, dykman1994large, tuanase2012exchange}. 
If a singular behavior of $b_c(r)$ is concluded, 
the two limiting cases reported in this Letter
could characterize the different phases. 
This is left as future work.

The authors thank Andreas Dechant, Namiko Mitarai, Naoko Nakagawa, Masato Itami and Ikumi Kobayashi for helpful discussions.
The work of Y.Y. was supported by JST, the establishment of university fellowships towards
the creation of science technology innovation, Grant Number JPMJFS2123.
This work was supported by JSPS KAKENHI (Grant Numbers JP19H05795 and JP22H01144).





%


\clearpage
\onecolumngrid

\setcounter{equation}{0}
\renewcommand{\theequation}{S\arabic{equation}}

\setcounter{figure}{0}
\renewcommand{\thefigure}{S\arabic{figure}}

\begin{center}
{\large \bf Supplemental Material:  \protect \\ 
Phase coexistence in a weakly stochastic reaction-diffusion system}\\
\vspace*{0.3cm}
Yusuke Yanagisawa and Shin-ichi Sasa
\end{center}

This Supplemental Material consists of three sections. 
In Sec. \ref{derivations}, we derive the formulas presented in the main text.
In Sec. \ref{numerics}, we show some numerical results supporting arguments in the main text.
Finally, in Sec. \ref{test}, we discuss a possibility of testing our theory experimentally.

\section{Derivation of formulas} \label{derivations}

\subsection{Derivation of \eqref{descrete_phi}}\label{der-12}
Let $\mathbb{Z}_{\geq m}$ denote the set of integers greater than 
or equal to an integer $m$. 
Although particle numbers satisfy $X_i\in \mathbb{Z}_{\geq0}$
for all $i$, for later convenience, we assume $X_i\in \mathbb{Z}_{\geq-1}$
and set $P_t(\vb*{X})\equiv 0$ if there exists $i$ such that $X_i=-1$. 

To simplify the notation, we introduce operators $R_i,L_i : \mathbb{Z}_{\geq0}^{2N+1} \to \mathbb{Z}_{\geq-1}^{2N+1}$ as
\begin{align}
    R_i \vb*{X} &= \vb*{X}+\vb*{1}_i-\vb*{1}_{i+1} \quad (i\in[-N,N-1]), \\ 
    L_i \vb*{X} &= \vb*{X}-\vb*{1}_{i-1}+\vb*{1}_i \quad (i\in[-N+1,N]),
\end{align}
where $\vb*{1}_i\equiv (0,\cdots,0,1,0,\cdots,0)$ is a unit vector whose $i$-th component is one.
The probability distribution $P_t(\vb*{X})$ of the state $\vb*{X}\in\mathbb{Z}_{\geq -1}^{2N+1}$
obeys a master equation
\begin{align}
        \partial_t P_t(\vb*{X})
        &=\sum_{i=-N}^N (W_+(X_i-1)P_t(\vb*{X}-\vb*{1}_i)-W_+(X_i)P_t(\vb*{X})) 
        + \sum_{i=-N}^N (W_-(X_i+1)P_t(\vb*{X}+\vb*{1}_i)-W_-(X_i)P_t(\vb*{X})) \nonumber \\
        &\quad + r\sum_{i=-N}^{N-1}((X_i+1)P_t(R_i\vb*{X})-X_iP_t(\vb*{X})) +r\sum_{i=-N+1}^{N}((X_i+1)P_t(L_i\vb*{X})-X_iP_t(\vb*{X})),
\end{align}
where the transition rates for the Schl\"{o}gl model are given by
\begin{align}
    W_+(X) &= k_{+1}a\Omega + k_{-2}bX(X-1)/\Omega, \\
    W_-(X) &= k_{-1}X + k_{+2}X(X-1)(X-2)/\Omega^2.
\end{align}
Let $M_i(t)$ be the expectation value of the number of particles in the $i$-th vessel.
It is expressed as
\begin{equation}
M_i(t) \equiv \sum_{\vb*{X}\in \mathbb{Z}_{\geq 0}^{2N+1}} X_i P_t(\vb*{X}) = \sum_{X_{-N}=0}^\infty\cdots\sum_{X_N=0}^\infty X_i P_t(\vb*{X}).
\end{equation}
We then obtain 
\begin{align}
  \partial_tM_i(t) = \sum_{\vb*{X}\in \mathbb{Z}_{\geq0}^{2N+1}}
  W_+(X_i)P_t(\vb*{X}) -\sum_{\vb*{X}\in \mathbb{Z}_{\geq0}^{2N+1}} W_-(X_i)P_t(\vb*{X}) 
        + r(M_{i+1}(t)-2M_i(t)+M_{i-1}(t))
    \label{descrete_Mi}
\end{align}
with the boundary conditions
$M_{-(N+1)}(t) = M_{-N}(t)$ and  $M_{N+1}(t) = M_N(t)$.
In the limit $\Omega\to\infty$, the first and second terms in \eqref{descrete_Mi} become
\begin{align}
      &\quad \lim_{\Omega \to \infty}\frac{1}{\Omega}
      \qty(\sum_{\vb*{X}\in \mathbb{Z}_{\geq0}^{2N+1}}
      W_+(X_i)P_t(\vb*{X}) -\sum_{\vb*{X}\in \mathbb{Z}_{\geq0}^{2N+1}}
      W_-(X_i)P_t(\vb*{X})) \nonumber \\
      &= k_{+1}a +k_{-2}b \qty(\phi_i(t))^2
      -\qty(k_{-1}\phi_i(t) +k_{+2}\qty(\phi_i(t))^3),
\end{align}
where $\phi_i(t) \equiv \lim_{\Omega \to \infty}  M_i(t)/\Omega$.
Therefore, (\ref{descrete_Mi}) becomes 
\begin{align}
        \partial_t\phi_i = -k_{+2}\phi_i^3+k_{-2}b\phi_i^2-k_{-1}\phi_i+k_{+1}a
        + r\qty(\phi_{i+1}-2\phi_i+\phi_{i-1}).
\end{align}
The boundary conditions are
$\phi_{-(N+1)}(t) = \phi_{-N}(t)$ and $\phi_{N+1}(t) = \phi_N(t)$.

\subsection{Phase coexistence condition for \eqref{RD_eq}}
Here, we derive the phase coexistence condition for the reaction-diffusion equation \eqref{RD_eq}.
A phase coexistence state $\phi(x)$ is defined as a spatially inhomogeneous steady state 
connecting two locally stable stationary concentrations.
Then, $\phi(x)$ satisfies
\begin{align}
    0 = -k_{+2}\phi^3+k_{-2}b\phi^2-k_{-1}\phi+k_{+1}a + D\pdv[2]{\phi}{x} = -\dv{V_{\mathrm{det}}(\phi)}{\phi} + D\pdv[2]{\phi}{x}.
    \label{phase_coexist}
\end{align}
Note that \eqref{phase_coexist} is regarded as Newton's equation 
for the position $\phi$ as a function of the fictitious time $x$, where the potential is given by $-V_{\mathrm{det}}(\phi)$.
In this analogy, the phase coexistence state is expressed as a heteroclinic orbit which joins the two maximum points of $-V_{\mathrm{det}}(\phi)$.
Because of the energy conservation, such a heteroclinic orbit exists only when
the potential takes the same values at the two maximum points.
Thus, the phase coexistence condition for the reaction-diffusion equation \eqref{RD_eq} is given by $b=b_c^{\mathrm{det}}$.

\subsection{Derivation of \eqref{J_PDE}}
In the ramped system, $b_i = b + i\Delta b/N \equiv b + \delta b_i$ is defined for the $i$-th vessel.
Because the transition rate $W_+$ of the ramped system depends on $b_i$, 
we explicitly write its dependence as $W_+(X_i;b_i)$ in this subsection. 
We thus write
\begin{align}
    W_+(X_i;b_i) = k_{+1}a\Omega + k_{-2}b_i X_i(X_i-1)/\Omega = \Omega w_+(\xi_i;b_i) + o(\Omega),
\end{align}
where 
\begin{align}
    w_+(\xi_i;b_i) = k_{+1}a+k_{-2}b_i\xi_i^2 = w_+(\xi_i;b) + O(\Delta b).
\end{align}
The stationary distribution of the system $P_{\mathrm{ss}}(\vb*{X})$ satisfies
\begin{align}
    0 &= \sum_{i=-N}^N (W_+(X_i-1;b_i)P_{\mathrm{ss}}(\vb*{X}-\vb*{1}_i)-W_+(X_i;b_i)P_{\mathrm{ss}}(\vb*{X})) \nonumber \\
    &\quad + \sum_{i=-N}^N (W_-(X_i+1)P_{\mathrm{ss}}(\vb*{X}+\vb*{1}_i)-W_-(X_i)P_{\mathrm{ss}}(\vb*{X})) \nonumber \\
    &\quad + r\sum_{i=-N}^{N-1}((X_i+1)P_{\mathrm{ss}}(R_i\vb*{X})-X_iP_{\mathrm{ss}}(\vb*{X})) +r\sum_{i=-N+1}^{N}((X_i+1)P_{\mathrm{ss}}(L_i\vb*{X})-X_iP_{\mathrm{ss}}(\vb*{X})).
    \label{master_eq_perturbative}
\end{align}
Next, we assume the asymptotic form of the stationary solution of the master equation as
\begin{align}
    P_{\mathrm{ss}}(\vb*{X}) = \exp\qty(-\Omega\tilde{V}_{\mathrm{sto}}(\vb*{\xi};b,\Delta b)+o(\Omega)),
    \label{ss_dist}
\end{align}
where we expand
\begin{align}
    \begin{split}
        \tilde{V}_{\mathrm{sto}}(\vb*{\xi};b,\Delta b) 
        &= \sum_{i=-N}^N V_{\mathrm{sto}}(\xi_i;b)
        +\sum_{i=-N}^N \partial_b V_{\mathrm{sto}}(\xi_i;b)\cdot\delta b_i +rJ(\vb*{\xi}) \\
        &\equiv \sum_{i=-N}^N V_{\mathrm{sto}}^{(i)}(\xi_i) +rJ(\vb*{\xi}), \label{V_tilde}
    \end{split}
\end{align}
which is the stochastic potential of the ramped system up to the order of $\Delta b$ and $r$.
Here, $V_{\mathrm{sto}}^{(i)}(\xi_i)$ satisfies the relation
\begin{align}
    \exp(\partial_{\xi_i}V_{\mathrm{sto}}^{(i)}(\xi_i)) = \frac{w_-(\xi_i)}{w_+(\xi_i;b_i)},
    \label{1box_ptl}
\end{align}
as shown in \eqref{def:spot} of the main text.
Substituting the stationary solution \eqref{ss_dist} into \eqref{master_eq_perturbative} 
and collecting the leading terms for large $\Omega$, 
we obtain 
\begin{align}
        0 &= \sum_{i=-N}^N w_+(\xi_i;b_i) \qty(\exp\qty(\partial_{\xi_i}V_{\mathrm{sto}}^{(i)}(\xi_i)+r\partial_{\xi_i}J)-1)
        + \sum_{i=-N}^N w_-(\xi_i) \qty(\exp\qty(-\partial_{\xi_i}V_{\mathrm{sto}}^{(i)}(\xi_i)-r\partial_{\xi_i}J)-1) \nonumber \\
        &\quad +r \sum_{i=-N}^{N-1} \xi_i \qty(\exp\qty(-\partial_{\xi_i}V_{\mathrm{sto}}^{(i)}(\xi_i)+\partial_{\xi_{i+1}}V_{\mathrm{sto}}^{(i+1)}(\xi_{i+1})
        -r\partial_{\xi_i}J +r\partial_{\xi_{i+1}}J)-1) \nonumber \\
        &\quad +r \sum_{i=-N+1}^N \xi_i \qty(\exp\qty(\partial_{\xi_{i-1}}V_{\mathrm{sto}}^{(i-1)}(\xi_{i-1})-\partial_{\xi_i}V_{\mathrm{sto}}^{(i)}(\xi_i)
        +r\partial_{\xi_{i-1}}J -r\partial_{\xi_i}J)-1).
\end{align}
By considering \eqref{1box_ptl}, we find that the equation becomes
\begin{align}
    \sum_{i=-N}^N (w_+(\xi_i;b)-w_-(\xi_i))\partial_{\xi_i}J
    = \sum_{i=-N}^{N-1} \qty[\xi_i (g_i(\vb*{\xi})-1) +\xi_{i+1} (g_i(\vb*{\xi})^{-1}-1) ] \label{J_PDE_suppl}
\end{align}
with
\begin{align}
    g_i(\vb*{\xi}) = \frac{w_+(\xi_i;b) w_-(\xi_{i+1})}{w_-(\xi_i)w_+(\xi_{i+1};b)}.
\end{align}
\eqref{J_PDE_suppl} corresponds to \eqref{J_PDE} in the main text.

\subsection{Solution $J(\vb*{\xi})$ to \eqref{J_PDE}}
\label{chara}
We numerically solve \eqref{J_PDE} in the main text by using the characteristic
curve method. We first introduce a fictitious time $s$  and define
$\xi_{i}(s)$ by 
\begin{align}
  \dv{\xi_i(s)}{s} = w_+(\xi_i) - w_-(\xi_i),
  \label{rate_eq_characteristic} 
\end{align} 
for $i\in [-N,N]$.
Using \eqref{J_PDE}, we find that $J(\vb*{\xi})$ satisfies 
\begin{align}
  \dv{J(\vb*{\xi}(s))}{s}
  &= \sum_{i=-N}^{N-1}
  \qty[\xi_i (g_i(\vb*{\xi})-1) +\xi_{i+1} (g_i(\vb*{\xi})^{-1}-1) ]
  \label{J_characteristic}.
\end{align}
For a given initial concentration profile $\vb*{\xi}^{\mathrm{ini}}$,
we can determine a stationary solution $\vb*{\xi}^{\mathrm{fin}}$ of 
\eqref{rate_eq_characteristic}. By integrating  \eqref{J_characteristic}
with respect to $s \in [0, \infty)$, we obtain the difference
$J(\vb*{\xi}^{\mathrm{fin}})-J(\vb*{\xi}^{\mathrm{ini}})$
for the trajectory from $\vb*{\xi}^{\mathrm{ini}}$ to $\vb*{\xi}^{\mathrm{fin}}$.

Here, \eqref{rate_eq_characteristic} is equivalent to 
the rate equation for the well-mixed Schl\"{o}gl model.
We set $\xi^{\mathrm{ini}}_i=\xi_0\pm\Delta\xi$, where $\xi_0$ is
the unstable stationary solution of \eqref{rate_eq_characteristic}
and $\Delta \xi$ satisfies $0  <  \Delta \xi \ll \xi_0$.
Then, as the initial concentration profile $\vb*{\xi}^{\mathrm{ini}}$, 
we choose that all components other than $\xi^{\mathrm{ini}}_i$ are
either $\xi_+$ or $\xi_-$, which are the stable stationary solution
of \eqref{rate_eq_characteristic}. In this case, the only $\xi_i$
evolves in time and eventually approaches $\xi_\pm$, respectively.
Using these trajectories, we successively calculate the difference
of $J(\vb*{\xi})$ between $2N+2$ configurations
$\vb*{\xi}^{(0)},\vb*{\xi}^{(1)},\cdots,\vb*{\xi}^{(2N+1)}$,
where for any integer $k$ satisfying $0\leq k \leq 2N+1$,
$\xi^{(k)}_i=\xi_-$ for $-N \le i \le N-k$ and
$\xi^{(k)}_i=\xi_+$ for $N-k+1 \le i \le N$. 
Thus, we obtain $J(\vb*{\xi}^{(k)})$ up to an additive constant.

\section{Supplemental data}
\label{numerics}

\subsection{Details for Fig. \ref{bc-Omega_relation}}

\begin{figure}[t]
    \begin{minipage}[t]{0.48\hsize}
    \center
    \includegraphics[width=7cm]{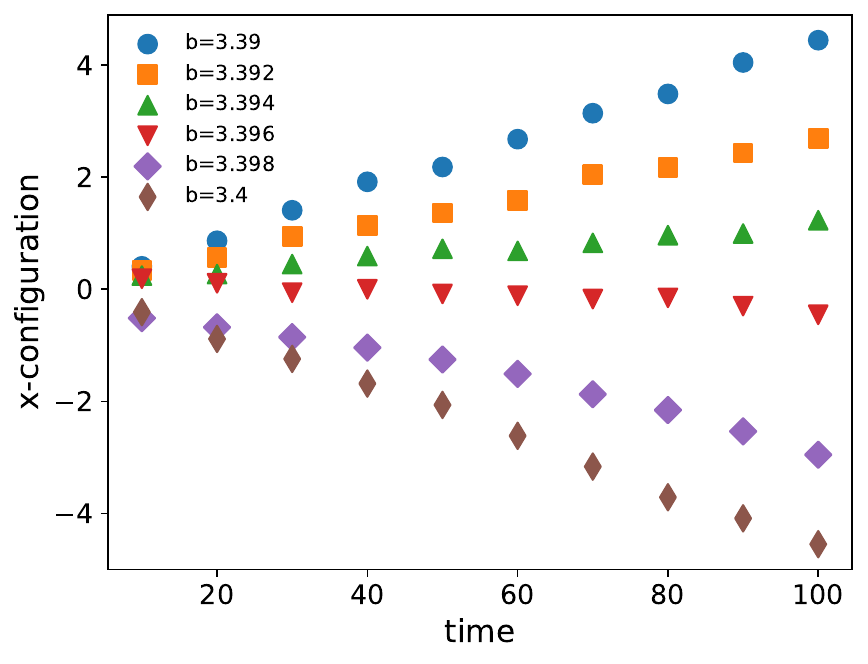}
    \caption{$\langle\hat{z}(t)\rangle$ for $r=5.0$ and $\Omega=400$.}
    \label{time_evol_interface}
    \end{minipage}
    \hfill
    \begin{minipage}[t]{0.48\hsize}
    \center
    \includegraphics[width=7cm]{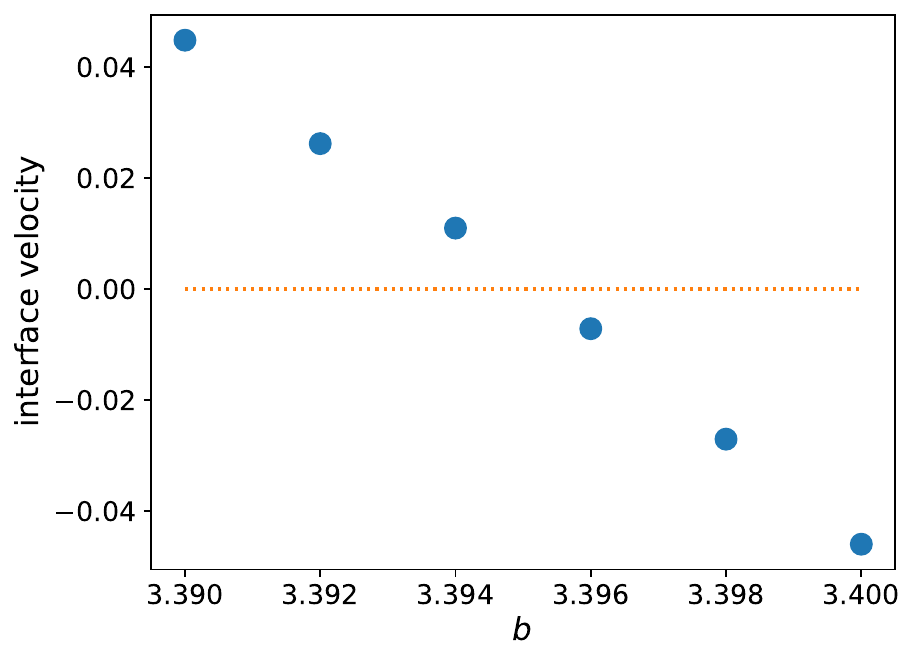}
    \caption{$b$ dependence of $v(b)$ for $r=5.0$ and $\Omega=400$.
    The dotted line shows $v=0$.}
    \label{interface_vel_vs_b}
    \end{minipage}
\end{figure}

Figure \ref{bc-Omega_relation} of the main text shows the phase coexistence condition $b_c(\Omega)$ 
that are numerically obtained for $r=5.0$.
In this subsection, we explain the method of numerical calculation in detail.

First, we need to determine the interface position $\hat{z}(t)$ at time $t$ in numerical simulations of the stochastic reaction-diffusion system.
We define $\hat{z}(t)$ as the value of the $x$-coordinate at which the concentration profile $\vb*{X}(t)/\Omega$ crosses the reference concentration $\xi_{\mathrm{ref}}\equiv(\xi_++\xi_-)/2$.
Explicitly, when $X_j(t)$ and $X_{j+1}(t)$ satisfy
$X_j(t)<\Omega\xi_{\mathrm{ref}}$ and $X_{j+1}(t)>\Omega\xi_{\mathrm{ref}}$,
the interface position $\hat{z}(t)$ is given by
\begin{align}
    \hat{z}(t) \equiv j\Delta x + \frac{\Omega\xi_{\mathrm{ref}}-X_j(t)}{X_{j+1}(t)-X_j(t)}\Delta x.
\end{align}

\begin{figure}
    \centering
    \includegraphics[width=\linewidth]{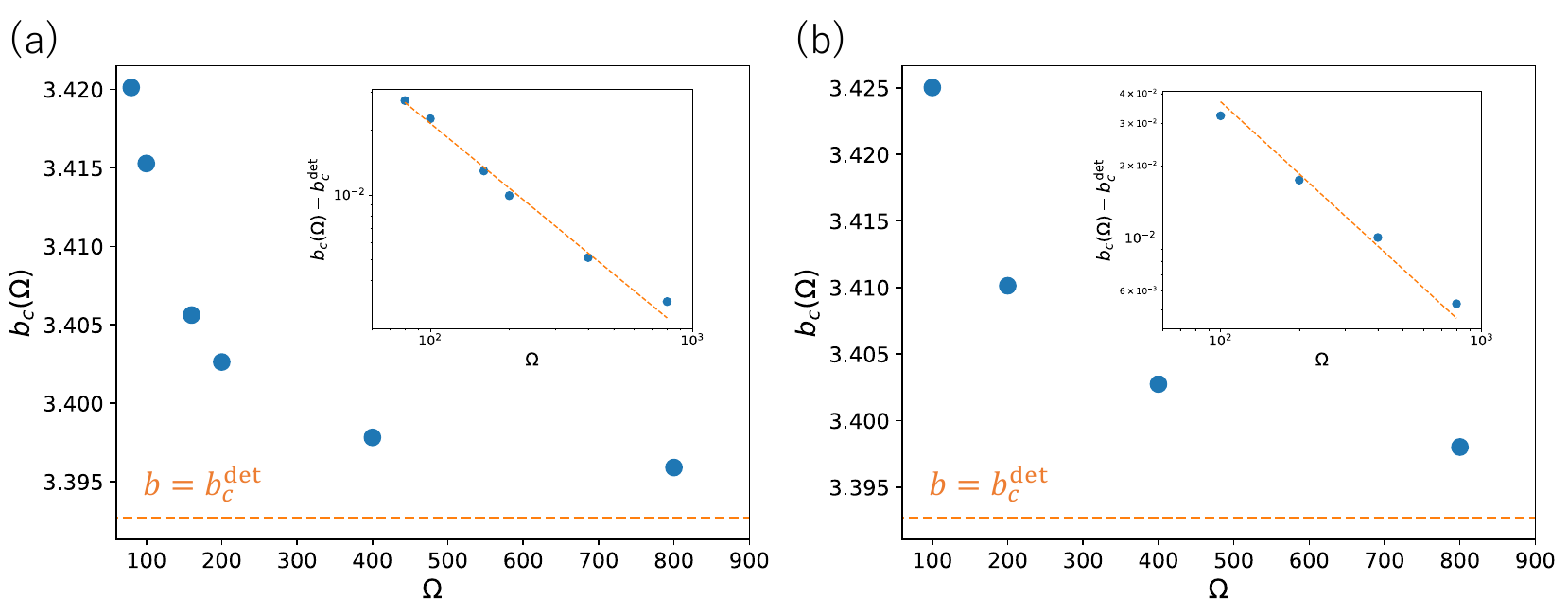}
    \caption{$\Omega$ dependence of $b_c(\Omega)$ for (a) $r=1.0$ and (b) $r=0.5$. The inset
    shows a log-log plot with the guide line $b_c(\Omega)-b_c^{\mathrm{det}} = \alpha/\Omega$, where (a) $\alpha=2.15$ and (b) $\alpha=3.7$.}
    \label{b_c(Omega)_vs_Omega_r1_0_0_5}
\end{figure}

Next, for several values of $b$, we calculate the average of the interface position $\langle\hat{z}(t)\rangle$ at time $t$.
Figure \ref{time_evol_interface} is the example of $\langle\hat{z}(t)\rangle$ for $r=5.0$ and $\Omega=400$, where $\langle\cdot\rangle$ is estimated using $100$ samples.
From the data of $\langle\hat{z}(t)\rangle$, we estimate the interface velocity $\hat{v}(b)$ as 
\begin{align}
    v(b) = \frac{\langle\hat{z}(t_1)\rangle - \langle\hat{z}(t_0)\rangle}{t_1-t_0}, 
\end{align}
where $t_0$ is much larger than the relaxation time $\kappa_{\pm}^{-1}$,
and $t_1-t_0$ is large enough to  detect the average velocity.
We set $t_0=10$ and $t_1=100$.
Figure \ref{interface_vel_vs_b} shows the $b$ dependence of $v(b)$ for $r=5.0$ and $\Omega=400$.
We proceed to determine $b_c(\Omega)$ from the data of $v(b)$.
Let $db$ be a step width of $b$ used in numerical simulations.
We then find a special $b_o$ such that $v(b_o)>0$ and $v(b_o+db)<0$.
Using this $b_o$, we obtain $b_c(\Omega)$ by the formula
\begin{align}
    b_c(\Omega) \equiv b_o + \frac{v(b_o)}{v(b_o)-v(b_o+db)}db.
\end{align}
The results for several values of $\Omega$ are shown in Fig. \ref{bc-Omega_relation} of the main text.
Similar behaviors are observed for $r=1.0$ and $0.5$, as shown in Fig. \ref{b_c(Omega)_vs_Omega_r1_0_0_5}.
All these results indicate that $b_c(\Omega) \to b_c^{\mathrm{det}}$ as $\Omega\to\infty$.

\subsection{The asymptotic behavior of $\Delta b_*$}
\begin{figure}
    \centering
    \includegraphics[width=0.5\linewidth]{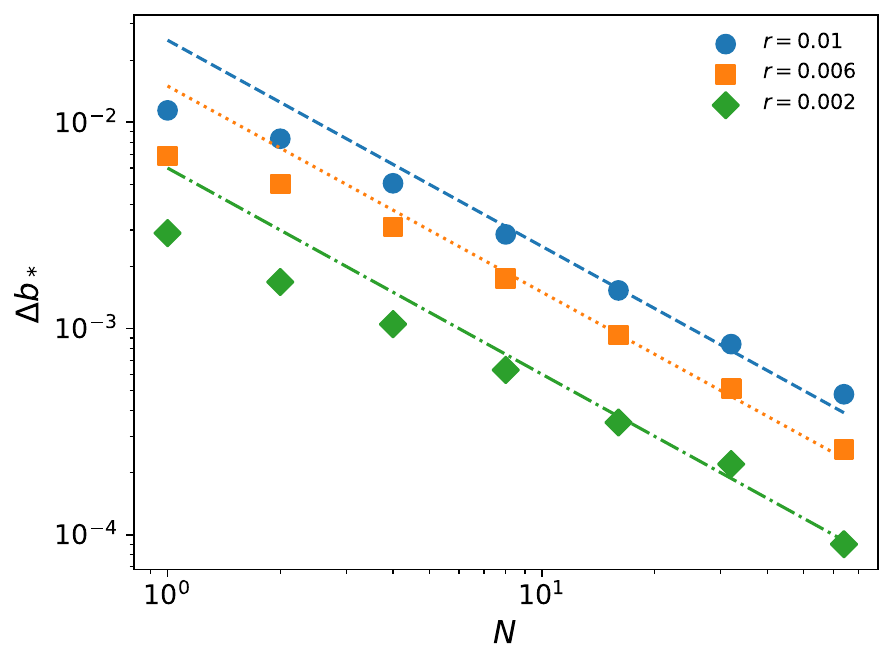}
    \caption{$N$ dependence of $\Delta b_*$ for $r=0.01, 0.006$ and $0.002$.
    The guide lines show $\Delta b_*=\alpha/N$, where $\alpha=0.025$ for the blue dashed line, $\alpha=0.015$ for the orange dotted line and $\alpha=0.006$ for the green dash-dot line.}
    \label{delbstar-N_relation}
\end{figure}

In the main text, we introduce $\Delta b_*$ which is necessary to observe the phase coexistence state for systems with finite $N$.
In Fig. \ref{delbstar-N_relation}, we show that $\Delta b_*$ decreases with respect to $N$.
More precisely, $\Delta b_*$ approaches $0$ following the law $\Delta b_*\simeq 1/N$ for large $N$.
This means that we do not need the parameter ramping to observe the phase coexistence in the limit $N\to\infty$.

\section{Experimental testing of our theory}
\label{test}
In this section, we discuss a possibility to observe phase coexistence in the low-hopping-rate regime.
The Kuramoto length for reaction-diffusion systems is estimated from diffusion constants and reaction rate constants 
using data in Refs. \cite{sneppen2014models_v2,alon2006an_v2,murray2002mathematical_v2}.
If the relevant chemical spiecies is a protein, the typical diffusion constant is $0.1\sim 100\;\si{\micro\meter^2\s^{-1}}$ 
and the typical reaction rate constant is about $100\;\si{\s}$.
The Kuramoto length for such systems is $1\sim 100\;\si{\micro\meter}$.
As another example, if the relevant chemical spiecies is a small molecule, the typical diffusion constant is $10\sim 10000\;\si{\micro\meter^2\s^{-1}}$
and the typical reaction rate constant is about $10^{-2}\;\si{\s}$.
The Kuramoto length is then $0.1\sim 10\;\si{\micro\meter}$.
Here, we consider a reaction-diffusion system
by using an assemble of bacteria \cite{karig2018stochastic_v2}.
Suppose that there are bacteria in which a bistable chemical reaction occurs.
These bacteria can be considered as reaction vessels.
Chemical spiecies inside each bacterium can be released to the environment
and they can also be absorbed by surrounding bacteria.
The system can be identified as a stochastic reaction-diffusion system that we consider.
Because the typical body length of bacteria is $2\;\si{\micro\meter}$,
one may construct a stochastic reaction-diffusion system in the low-hopping-rate regime.

\end{document}